\documentstyle[preprint,aps]{revtex}
\begin{document}
\draft
\preprint{}
\title{Magnetic Charge as a ``Hidden'' Gauge Symmetry}
\author{D. Singleton}
\address{Department of Physics, University of Virginia, 
Charlottesville, VA 22901}
\date{\today}
\maketitle
\begin{abstract}
A theory containing both electric and magnetic charges is formulated
using two vector potentials, $A ^{\mu}$ and $C ^{\mu}$. This has 
the aesthetic advantage of treating electric and magnetic charge both
as gauge symmetries, but it has the experimental disadvantage 
of introducing a second massless gauge boson (the ``magnetic'' photon) 
which is not observed. This problem is dealt with by using the 
Higgs mechanism to give a mass to one of the gauge bosons while the other
remains massless. This effectively ``hides'' the magnetic charge, and the
symmetry associated with it, when one is at an energy scale far enough
removed from the scale of the symmetry breaking.
\end{abstract}
\pacs{PACS numbers: 14.80.H, 11.10.E, 11.15.E}
\newpage
\narrowtext
\section{Introduction}
Since the seminal work of Dirac \cite{dirac} magnetic 
monopoles have excited much theoretical interest, but there has been
no confirmed experimental evidence of their existence up to the 
present. Dirac's formulation requires the introduction of a singular vector 
potential so that the definition, ${\bf B} = \nabla \times {\bf A}$,
may be used, while still having $\nabla \cdot {\bf B} = \rho _m$,
where $\rho _m$ is the magnetic charge density. The vector potential,
${\bf A}$, is singular along a line which runs from the magnetic charge
off to spatial infinity. By requiring that the string singularity have
no physical effect (i.e. the wavefunction of a charged particle must 
vanish along it) one arrives at Dirac's condition for the
quantization of electric charge. A fiber-bundle formulation of 
magnetic monopoles has been given by Wu and Yang \cite{yang}
which avoids the need for a singular vector potential,
but defines the vector potential differently in two different regions
surrounding the magnetic charge. The two vector potentials are related
by a gauge transformation, and requiring that the gauge transformation
function be single-valued yields the Dirac quantization condition again.

In this article we wish to present a different formulation of a magnetic
charge based upon the gauge principle. Electric charge is a gauge 
charge which is coupled to a gauge field, the photon, by replacing 
the ordinary derivative with the gauge-covariant derivative in the
Lagrangian. In electrodynamics the gauge field corresponds to the 
four-vector potential, $A _{\mu}$. Since the generalized Maxwell equations
with electric and magnetic charge appear symmetric between the two
types of charges and currents one might ask if it is possible to treat
magnetic charge, like electric charge, as a gauge symmetry. 
The gauge principle then implies that there must be a second 
massless gauge boson corresponding to magnetic charge. In the 
next section we will show that Maxwell's equations with electric and 
magnetic charge naturally have room for two four-vector
potentials, $A_{\mu}$ and $C_{\mu}$. One is then left with two massless 
photons while only one massless photon is known to exist experimentally.
This difficulty can be overcome through 
the use of the Higgs mechanism \cite{higgs}
which allows one to make gauge bosons massive while not
violating the gauge invariance of the theory. This allows the charge
which is coupled to the massive gauge boson to remain ``hidden'' as
long as one is not too near the energy scale of the symmetry breaking.

There have been previous attempts to construct a field theory of 
magnetic charges in terms of a pair of four-potentials \cite{cabbibo} 
\cite{zwang}, but the second vector potential has always been 
somewhat problematic since then there are apparently too many degrees of 
freedom. Spontaneous symmetry breaking allows one to deal with these extra 
degrees of freedom in a natural way. 

\section{Generalized Maxwell Equations and Duality}

The generalized Maxwell equations in the presence of electric 
and magnetic charges and currents are \cite{jackson}
\begin{eqnarray}
\nabla \cdot {\bf E} = \rho _e  \; \; \; \; \;  \nabla \times {\bf B} =
{1 \over c}\left( {\partial {\bf E} \over \partial t} +
{\bf J} _e \right) \nonumber \\
\nabla \cdot {\bf B} = \rho _m  \; \; \; \; \;  - \nabla \times {\bf E} =
{1 \over c} \left( {\partial {\bf B} \over \partial t} +
{\bf J} _m \right)
\label{maxgen}
\end{eqnarray}
These equations are invariant under the following duality transformation
\begin{eqnarray}
{\bf E} \rightarrow {\bf E} cos \theta + {\bf B} sin \theta \nonumber \\
{\bf B} \rightarrow -{\bf E} sin \theta + {\bf B} cos \theta
\label{dual1}
\end{eqnarray}
\begin{eqnarray}
\rho _e \rightarrow \rho _e cos \theta + \rho _m sin \theta \; \; 
\; \; \;  \rho_m \rightarrow - \rho _e sin \theta + \rho _m cos \theta
\nonumber \\
{\bf J}_e \rightarrow {\bf J}_e cos \theta + {\bf J}_m sin \theta \; \; 
\; \; \; {\bf J}_m \rightarrow - {\bf J}_e sin \theta + {\bf J}_m cos \theta
\label{dual2}
\end{eqnarray}

Now one can introduce two four-vector potentials, $A ^{\mu} =
(\phi _e , {\bf A})$ and $C ^{\mu} = (\phi _m , {\bf C})$ and write
the ${\bf E}$ and ${\bf B}$ fields as
\begin{eqnarray}
{\bf E} = - \nabla \phi _e - {1 \over c}{\partial {\bf A} \over
\partial t} - \nabla \times {\bf C} \nonumber \\
{\bf B} = -\nabla \phi _m - {1 \over c}{\partial {\bf C} \over
\partial t} + \nabla \times {\bf A}
\label{ebpoten}
\end{eqnarray}
In electrodynamics with only electric charge $\bf {E}$ consists of only 
the first two terms, while $\bf {B}$ consists of only the last term. 
Taking $\phi _e$ as a scalar, and $\bf {A}$ as a vector under
spatial inversion leads to $\bf {E}$ being a vector and $\bf {B}$ being a
pseudovector under spatial inversion. In order for $\bf {E}$ to remain a
vector, and $\bf {B}$ to remain a pseudovector in Eq. (\ref{ebpoten}) 
$\phi _m$ must be a pseudoscalar and $\bf {C}$ a pseudovector. 
Using the gauge freedom which is possessed 
by the potentials one can chose, $A ^{\mu}$ and $C ^{\mu}$ such 
that they satisfy the Lorentz gauge condition
\begin{eqnarray}
\partial _{\mu} A ^{\mu} = {1 \over c}{\partial \phi _e \over
\partial t} + \nabla \cdot {\bf A} = 0 \nonumber \\
\partial _{\mu} C ^{\mu} = {1 \over c}{\partial \phi _m \over
\partial t} + \nabla \cdot {\bf C} = 0 
\label{gaugecon}
\end{eqnarray}
On substituting the expressions for {\bf E} and {\bf B}, in terms of
the potentials Eq. (\ref{ebpoten}), into the generalized Maxwell
equations Eq. (\ref{maxgen}), using the two Lorentz gauge conditions 
Eq. (\ref{gaugecon}), and applying some standard vector calculus
identities ($\nabla \cdot [\nabla \times {\bf a}] = 0$, $\nabla \times 
[\nabla \phi] = 0$, and $\nabla \times [\nabla \times {\bf a}] =
\nabla [ \nabla \cdot {\bf a}] - \nabla ^2 {\bf a}$) 
one arrives at the following alternative form for the equations
\begin{eqnarray}
\nabla ^2 \phi _e - {1 \over c^2}{\partial ^2 \phi _e \over \partial t^2}
= - \rho _e \; \; \; \; \;  \nabla ^2 \phi _m 
- {1 \over c^2}{\partial ^2 \phi _m \over \partial t^2}
= - \rho _m  \nonumber \\
\nabla ^2 {\bf A} - {1 \over c^2}{\partial ^2
{\bf A} \over \partial t^2} = -{1 \over c} {\bf J} _e 
\; \; \; \; \;  \nabla ^2 {\bf C} - {1 \over c^2}{\partial ^2
{\bf C} \over \partial t^2} = -{1 \over c} {\bf J} _m 
\label{maxwave}
\end{eqnarray}
From here on we will set $c = 1$. Since $\phi_m$ is a 
pseudoscalar and $\bf {C}$ a pseudovector Eq. (\ref{maxwave})
implies that the magnetic charge density, $\rho _m$, and 
magnetic current density, $\bf{J _m}$ are pseudoscalars 
and pseudovectors respectively. The significance of Eq. (\ref{maxwave})
is that it demonstrates that Maxwell's equations with both electric
and magnetic charge naturally allow for two four-vector potentials
(i.e. two ``photons''). It is 
desirable to cast our results to this point in covariant notation
in terms of the two four potentials, $A ^{\mu}$ and $C ^{\mu}$, in
order to be able to obtain a Lagrangian density for the generalized
Maxwell equations. Defining two field strength tensors in terms of
the two four-vector potentials
\begin{eqnarray}
F ^{\mu \nu} = \partial ^{\mu} A ^{\nu} - \partial ^{\nu} A ^{\mu}
\nonumber \\
G ^{\mu \nu} = \partial ^{\mu} C ^{\nu} - \partial ^{\nu} C ^{\mu}
\end{eqnarray}
One can write the Maxwell equations with magnetic charge,
Eq. (\ref{maxwave}), in the following covariant form
\begin{eqnarray}
\partial _{\mu} F ^{\mu \nu} = \partial _{\mu} \partial ^{\mu} A^{\nu}
 = J _e ^{\nu} \nonumber \\
\partial _{\mu} G ^{\mu \nu} = \partial _{\mu} \partial ^{\mu} C^{\nu}
 = J _m ^{\nu} 
\label{maxwave1}
\end{eqnarray}
where the Lorentz gauge of Eq. (\ref{gaugecon}) has been used.

The duality transformations of Eqs. (\ref{dual1}) ,(\ref{dual2}) can now
be written in terms of the four-potentials and four-currents
\begin{eqnarray}
A ^{\mu} \rightarrow A ^{\mu} cos \theta + C ^{\mu} sin \theta \nonumber \\
C ^{\mu} \rightarrow - A ^{\mu} sin \theta + C ^{\mu} cos \theta
\label{dualc}
\end{eqnarray}
\begin{eqnarray}
J_e ^{\mu} \rightarrow J_e ^{\mu} cos \theta + J_m ^{\mu} sin \theta
\nonumber \\
J_m ^{\mu} \rightarrow - J_e ^{\mu} sin \theta + J_m ^{\mu} cos \theta
\label{dualc2}
\end{eqnarray}
Finally the {\bf E} and {\bf B} fields of Eq. (\ref{ebpoten}) can be 
written in terms of the field-strength tensors as
\begin{eqnarray}
E_i = F^{i 0} + {1 \over 2} \epsilon ^{ijk} G_{jk} =
F^{i 0} - {\cal G}^{i 0} \nonumber \\
B_i = G^{i 0} - {1 \over 2} \epsilon ^{ijk} F_{jk} =
G^{i 0} + {\cal F}^{i 0}
\label{ebcov}
\end{eqnarray}
where
\begin{eqnarray}
{\cal F} ^{\mu \nu} = {1 \over 2} \epsilon ^{\mu \nu \alpha \beta}
F _{\alpha \beta} \nonumber \\
{\cal G} ^{\mu \nu} = {1 \over 2} \epsilon ^{\mu \nu \alpha \beta}
G _{\alpha \beta}
\end{eqnarray}
$\epsilon^{\mu \nu \rho \sigma}$ is the totally antisymmetric fourth rank 
Levi-Civita tensor where $\epsilon^{0123} = +1$ with even permutations 
of the indices giving $+1$ and odd permutations giving $-1$. ${\cal F}
^{\mu \nu}$ and ${\cal G} ^{\mu \nu}$ are the duals of $F ^{\mu
\nu}$ and $G ^{\mu \nu}$.
From Eq. (\ref{ebcov}) it looks as if two ``photons'' are contributing
to the {\bf E} and {\bf B} fields. However using spontaneous symmetry
breaking via a scalar field one of the ``photons'' is made massive
thus effectively reducing the above definitions to their usual form
of $E _i = F^{i 0}$ and $B _i = - {1 \over 2} \epsilon^{ijk} F_{jk}$ as
long as one is not too close to the energy scale of the breaking.

It is now straightforward to write down a Lagrangian density which
gives the Maxwell equations, Eq. (\ref{maxwave1}). These
equations are simply two copies of the same equation with the source of
one being an electric four-current, $J_e ^{\mu}$ and the source of the
other being a magnetic four-current, $J_m ^{\mu}$. Thus the most obvious
thing to do is to add a new kinetic term and a new source term for 
the four-potential, $C ^{\mu}$
\begin{equation}
{\cal L}_M = -{1 \over 4} F_{\mu \nu} F ^{\mu \nu} -{1 \over 4}
G_{\mu \nu} G ^{\mu \nu} -  J_e ^{\mu} A_{\mu}
-  J_m ^{\mu} C_{\mu}
\label{lagrange}
\end{equation}
Variation of this Lagrange density with respect to $A ^{\mu}$ and 
$C ^{\mu}$ yield the field equations, Eq. (\ref{maxwave1}),
in the Lorentz gauge. This Lagrangian is invariant under the 
duality transformation of Eqs. (\ref{dualc}) , (\ref{dualc2}). This 
property will be used in the next section. Also the Lagrangian is
a scalar despite the fact that it contains pseudo-quantities
($J_m ^{\mu}$ and $C_{\mu}$) since these quantities only occur
in combinations which result in scalars under parity.

One drawback to this simple extension of the usual
Maxwell Lagrangian is that it does not yield the expected
energy-momentum tensor in terms of the {\bf E} and {\bf B} fields
as defined in  Eq. (\ref{ebcov}). In particular one would expect
there to be cross terms between the field strength tensors $F ^{\mu \nu}$
and $G ^{\mu \nu}$ coming, for example, from $T ^{00} = {1 \over 2}
({\bf E} ^2 + {\bf B} ^2)$ with the use of Eq. (\ref{ebcov}). In
addition Eq. (\ref{lagrange}), while giving the correct Maxwell
equations, makes it appear as if the electric and magnetic four-currents
are completely decoupled from one another contrary to physical intuition 
that an electric charge and magnetic charge would interact with one
another. However the Lagrange density and energy-momentum tensor are not
unique. One is free to add any four-divergence to the Lagrangian without
changing the Maxwell equations, and one is also free to add any
four-divergence with the proper antisymmetry property \cite{gold}
without changing the conservation laws or integral quantities associated 
with the energy-momentum tensor. By adding $+{1 \over 8} \epsilon 
^{\mu \nu \rho \sigma} F_{\mu \nu} G_{\rho \sigma}$ to the Lagrangian 
one mixes the two ``photons'' and obtains the usual energy-momentum 
tensor with all the cross terms that are implied by Eq. 
(\ref{ebcov}).  Adding $\epsilon ^{\mu \nu \rho \sigma}
F _{\mu \nu} G_{\rho \sigma}$ to the Lagrangian does not change
the Maxwell equations, Eq. (\ref{maxwave1}), obtained from the Lagrangian 
since it is a total four-divergence by the antisymmetry property of 
$\epsilon ^{\mu \nu \rho \sigma}$. In this paper we will not explicitly
write this cross term  in the Lagrangian, since all we require for 
the present development is the minimal Lagrangian that yields 
the Maxwell equations.  

\section{The Scalar Field}

The Lagrangian formulation of Maxwell's equations as expressed in Eq.
(\ref{lagrange}) possesses the theoretically pleasing feature of
treating electric and magnetic charge symmetrically, but has the
experimentally displeasing feature of introducing a second massless
gauge boson. This second gauge boson (and the symmetry associated with
it) can be ``hidden'' using the Higgs mechanism. First one introduces
a complex scalar field, $\Phi$ which carries a scalar electric charge, 
$q_e$, and a pseudoscalar magnetic charge, $q_m$, and must be gauged
with respect to both the vector and pseudovector potentials
($A ^{\mu}$ and $C ^{\mu}$). The Lagrangian is
\begin{eqnarray}
{\cal L} _S&=&(\partial _{\mu} + i q_e A_{\mu} + i q_m C_{\mu}) \Phi ^{\ast}
(\partial ^{\mu} - i q_e A^{\mu} - i q_m C^{\mu}) \Phi  - V(\Phi ^2) 
\nonumber \\
&-&{1 \over 4} F_{\mu \nu} F^{\mu \nu} 
- {1 \over 4} G_{\mu \nu} G^{\mu \nu}
\label{lgscalar}
\end{eqnarray}
where $\Phi ^{\ast}$ is the complex conjugate of $\Phi$. The form of the 
gauge-covariant derivative in Eq. (\ref{lgscalar}) is that 
of Ref. \cite{zwang} here applied to a scalar field rather than a spinor 
field. The electric and magnetic four currents, $J_e ^{\mu}$ and
$J_m ^{\mu}$, of Eq. (\ref{lagrange}) can be written in terms of the scalar
fields as
\begin{eqnarray}
\label{curr}
J_e ^{\mu} = i q_e [ \Phi ^{\ast} (D^{\mu} \Phi) - \Phi (D^{\mu}
\Phi)^{\ast}] \nonumber \\
J_m ^{\mu} = i q_m [ \Phi ^{\ast} (D^{\mu} \Phi) - \Phi (D^{\mu}
\Phi)^{\ast}] 
\end{eqnarray}
where $D ^{\mu} = (\partial ^{\mu} - iq_e A^{\mu} -iq_m C^{\mu})$ is the
gauge-covariant derivative. Since $q_e$ and $q_m$ are scalar and 
pseudoscalar quantities respectively it follows from Eq. (\ref{curr}) that
$J_e ^{\mu}$ and $J_m ^{\mu}$ are a real four-current and a pseudo 
four-current respectively.
The gauge group of the above Lagrangian is $U(1) \times U(1)$ 
\cite{carmeli} , which is not a semi-simple group. This leads one to
suspect that there will be no quantization condition between electric
an magnetic charge. The potential, $V(\Phi ^2)$, contains the usual
mass and quartic self-interaction terms in order to develop a VEV
\begin{equation}
V(\Phi ^2) = m^2 (\Phi ^{\ast} \Phi) + \lambda (\Phi ^{\ast} \Phi) ^2
\end{equation}
The self-interaction coupling constant, $\lambda$, is taken to be
positive definite, and for $m ^2 <$ 0 the potential aquires a
vacuum expectation value of
\begin{equation}
\langle \Phi \rangle = {\sqrt{-m^2 \over 2 \lambda}}  
\equiv  {v \over \sqrt{2}}
\label{vev}
\end{equation}
Parametrizing the complex field in terms of real fields with the VEV
chosen to lie along the real component yields
\begin{eqnarray}
\Phi (x)&=&{1 \over \sqrt{2}} ( v + \eta (x) + i \zeta (x) ) 
\nonumber \\
&\approx& {1 \over \sqrt{2}}(v + \eta (x)) e^{(i \zeta (x) /v)}
\end{eqnarray}
Now using the gauge freedom of the Lagrangian in Eq. (\ref{lgscalar}),
one can transform the $\zeta (x)$ field away by making a gauge
transformation to the unitary gauge
\begin{eqnarray}
\label{unigauge}
\Phi ' (x)&=&e^{(-i \zeta (x) /v)} \Phi (x) = {1 \over \sqrt{2}}
(v + \eta (x)) \nonumber \\
A_{\mu} '(x)&=&A_{\mu} (x) - {1 \over 2 q_e v} 
\partial _{\mu} \zeta (x) \\
C_{\mu} '(x)&=&C_{\mu} (x) - 
{1 \over 2 q_m v} \partial _{\mu} \zeta (x) \nonumber
\end{eqnarray}
${\cal L} _S$ is invariant under these transformations. Substituting these 
unitary gauge fields into Eq. (\ref{lgscalar}) yields the following result
\begin{equation}
{\cal L}_S = {\cal L}_0 + {\cal L}_I
\end{equation}
where ${\cal L}_0$ contains the kinetic energy and mass terms
\begin{eqnarray}
\label{kinlgs}
{\cal L}_0&=&{1 \over 2} (\partial _{\mu} \eta) (\partial ^{\mu} \eta)
+ {1 \over 2} \left( 2 m^2 \right) \eta ^2 
- {1 \over 4}F '_{\mu \nu}
{F'} ^{\mu \nu} - {1 \over 4}G' _{\mu \nu}{G'} ^{\mu \nu} \nonumber \\
&+&{1 \over 2}v^2 
( q_e ^2 A'_{\mu} {A'}^{\mu} + q_m ^2 C'_{\mu} {C'}^{\mu}
+ 2 q_e q_m A'_{\mu} {C'}^{\mu})
\end{eqnarray}
where $F'_{\mu \nu}$ and $G'_{\mu \nu}$ indicate that the field
strength tensors are in terms of the gauge transformed fields
$A'_{\mu}$ and $C'_{\mu}$. Note that both the field-strength tensors
and therefore the equations of motion derived from them, are invariant
under the gauge transformations of Eq. (\ref{unigauge}). The interaction
terms of the Lagrangian, ${\cal L} _I$, are
\begin{eqnarray}
\label{intlgs}
{\cal L}_I&=&- \lambda v \eta ^3 - {\lambda \over 4} \eta ^4 
+v \eta (q_e ^2 A'_{\mu} A'^{\mu} + q_m^2 C'_{\mu} C'^{\mu}
+2 q_e q_m A'_{\mu} C'^{\mu}) \nonumber \\ 
&+&{\eta ^2 \over 2} (q_e^2 A'_{\mu} A'^{\mu}
+ q_m^2 C'_{\mu} C'^{\mu} + 2 q_e q_m A'_{\mu} C'^{\mu})
\end{eqnarray}
A disturbing feature in both Eq. (\ref{kinlgs}) and Eq. (\ref{intlgs})
is the presence of the cross terms between $A'_{\mu}$ and $C'_{\mu}$,
which makes finding the mass spectrum of the gauge bosons after
symmetry breaking complicated. Also it appears that both gauge bosons
have become massive even though only one scalar degree of freedom
has been absorbed. However one can now use the freedom of the duality
transformation, Eq. (\ref{dualc}), to diagonalize away the cross terms
$A'_{\mu} C'^{\mu}$ in both ${\cal L}_0$ and ${\cal L} _I$. One can
write the gauge boson mass terms and gauge boson-scalar interaction terms
in the form of  $2 \times 2$ matrices
\begin{eqnarray}
K \; (A'_{\mu} \; C'_{\mu})
\left(
\begin{array}{cc}
q_e^2 &q_e q_m \\
q_e q_m &q_m^2 
\end{array} \right)
\left(
\begin{array}{c}
A'^{\mu} \\
C'^{\mu}
\end{array} \right)
\end{eqnarray}
where $K = {v^2 \over 2} , v , {1 \over 2}$ for the gauge
boson mass, tri-linear interaction, and quartic interaction terms
respectively. Now applying a duality
transformation as in Eq. (\ref{dualc}) with $cos \theta =  q_m / 
\sqrt{q_e^2 + q_m^2}$ and $sin \theta =  q_e / \sqrt{q_e^2 + q_m^2}$ 
the mass matrix and interaction matrices are diagonalized
\begin{eqnarray}
K \; (A''_{\mu} \; C''_{\mu})
\left(
\begin{array}{cc}
0 &0 \\
0 &(q_e^2 + q_m^2) 
\end{array} \right)
\left(
\begin{array}{c}
A''^{\mu} \\
C''^{\mu}
\end{array} \right)
\end{eqnarray}
The double primes indicate that these are the duality rotated fields which
are obtained from $A'_{\mu}$ and $C' _{\mu}$ which are themselves the gauge
transformed fields of the original $A_{\mu}$ and $C_{\mu}$. However, as
the Lagranian is invariant under both the gauge and duality transformations 
we will from here on drop the primes. This diagonalization of the mass 
matrix by a duality rotation is the same procedure that is used in the
Standard Model \cite{salam} to obtain the
physical mass specturm of the electroweak
gauge bosons. One potential problem is the 
cross term ${1 \over 8} \epsilon ^{\mu \nu \rho \sigma} F_{\mu \nu} 
G_{\rho \sigma}$ which is gauge invariant, but is not invariant under 
the duality rotation. Since it is a total divergence, with no effect 
on the field equations, it could be added to the Lagrangian at any point 
({\em i.e.} it could be added to the final Lagrangian, Eq. (\ref{flagrange}),
below). Alternatively, one could add it to the original
Lagrangian, Eq. (\ref{lgscalar}), with a coefficient in front, chosen
so that after the duality rotation it would become ${1 \over 8} \epsilon 
^{\mu \nu \rho \sigma} F_{\mu \nu} G_{\rho \sigma}$. The original particle
spectrum of two massless gauge bosons and two scalar fields has, through
the Higgs mechanism, become one massive scalar field, one massless gauge
field and one massive gauge field. Writing out the final form of the
Lagrangian gives
\begin{eqnarray}
{\cal L}_S &=& {1 \over 2} \left( \partial _{\mu} \eta \right) 
\left( \partial ^{\mu} \eta \right)
+ {1 \over 2} \left(2 m^2 \right) \eta ^2 
-{1 \over 4} F_{\mu \nu} F^{\mu \nu}
-{1 \over 4} G_{\mu \nu} G^{\mu \nu} \nonumber \\
&+&{1 \over 2} g^2 v^2 C_{\mu} C^{\mu} + g^2 v C_{\mu} C^{\mu} \eta
+{1 \over 4} g^2 C_{\mu} C^{\mu} \eta ^2 - \lambda v \eta ^3 
-{\lambda \over 4} \eta ^4
\label{flagrange}
\end{eqnarray}
where, $g = \sqrt{q_e^2 + q_m^2}$, is the coupling strength 
of the scalar field to the gauge boson $C_{\mu}$. The scalar field 
$\eta (x)$ can be said to carry a magnetic charge of strength $g$, whose
associated symmetry is broken leading to the gauge boson connected with
it, $C_{\mu}$, having a mass, $m_C = g v$. Therefore the scalar, $\eta (x)$,
will have a Yukawa field rather than a Coulomb field surrounding it, 
and the magnetic charge will not be detectable unless one probes 
down to distances of the order of $r = {1 \over m_C}$. Notice that 
$A_{\mu}$ is now completely decoupled from the scalar field, $\eta (x)$,
and that the Lagrangian of Eq. (\ref{flagrange}) only has one charge, $g$.
This corresponds to the well known result that if all particles in a 
theory have the same ratio of electric to magnetic charge it is always
possible to use the duality rotation of Eq. (\ref{dualc2}) to rotate
away one of the charges. In order to make both electric and magnetic
charges play a non-trivial role one could introduce a second 
complex scalar field, $\Phi _2 (x)$, with no self interaction, and 
whose original couplings, $q_e'$ and $q_m'$, are different from $q_e$ 
and $q_m$. $\Phi _2 (x)$ would have to undergo a phase rotation, in
conjunction with Eq. (\ref{unigauge}), in order to allow its Lagrangian
to remain invariant. (This phase transformation would depend on $q_e$, $q_m$,
$q_e '$, $q_m '$, $v$, and $\zeta (x)$, but the explicit expression is not 
given since we will not pursue this point in detail here). On performing
the duality rotation the gauge boson - $\Phi _2 (x)$ interaction matrices
would in general include couplings between both components of $\Phi _2$ and
both gauge bosons. Thus $\Phi _2 (x)$ would represent two real scalar fields
carrying both electric and broken magnetic charge, whose values would
depend on the arbitrary values $q_e '$ and $q_m '$.

\section{Discussion and Conclusions}
In this paper we have attempted to give a formulation of magnetic 
charge as a gauge symmetry which is ``hidden'' through
spontaneous symmetry breaking. This was done through the 
introduction of two gauge potentials, $A_{\mu}$ and $C_{\mu}$. The
presence of two massless gauge bosons in the Lagrangian was resolved
by introducing a complex scalar field, $\Phi (x)$, and using the Higgs
mechanism to give a mass to one of the gauge bosons while leaving the
other one massless. Thus the existence of the massive gauge boson and
the symmetry associated with it was hidden as long as one was at an 
energy scale less than the mass of the gauge boson, $m_C$.

Although this formulation of magnetic charge appears to be very
different from Dirac's  approach there is a certain 
correspondence : In Dirac's formulation the string is an extra, non-local 
degree of freedom which is ``hidden'' through the Dirac quantization
condition; in the present formulation the ``magnetic'' photon is an extra, 
local degree of freedom which is ``hidden'' through the Higgs mechanism.
One difference between this formulation and previous formulations, which
involved either singular potentials \cite{dirac} or
gauge potentials which are defined differently over different domains,
\cite{yang} is that we do not obtain a charge quantization condition.
However both Grand Unified gauge theories and Kaluza-Klein theories
give alternative explanations for the quantization of charge, so
this loss of an explanation for charge quantization might not be so
unpleasant. The broken charge ($g = \sqrt{q_e^2 + q_m^2}$) that is carried
by $\eta$ depends only on the unconstrained (at the classical level) 
values of $q_e$ and $q_m$. 

There is much arbitrariness in this formulation of a magnetic charge :
the charge $g$, the mass of $\eta$, the mass of
the gauge boson,  and the self coupling of $\eta$
are all unspecified. The aim of this paper was not to construct a
fully realistic model of a physical monopole, but to give a treatment
where magnetic charge is treated, initially, exactly as electric charge,
and then provide an explanation for the non-observation of the symmetry
associated with the magnetic charge. An interesting possibility of this
formulation is to use it to construct fermions out of bound states
of scalar particles carrying a broken magnetic charge and scalar particles
carrying an unbroken electric charge. If the bound state were bound to
a radius smaller than $r = {1 \over m_C}$ then the electric charge would
``see'' most of the broken magnetic charge. 
There would be an angular momentum 
associated with the bound state \cite{saha} which would depend on $g$,
$m_C$, the electric charge, and the size of the bound state. (For a system
with an unbroken magnetic charge and electric charge the angular momentum
is independent of the distance between the two charges, but with a broken
magnetic charge the closer the electric charge got the more magnetic charge
it would ``see''). The gauge couplings and symmetry breaking could then be
adjusted so as to let one interpret the angular momentum of the 
charge-monopole system as the spin of the bound state. Finally it has 
been shown that a charge-monopole system obeys Fermi-Dirac statistics 
\cite{goldhaber}. It is plausible to postulate that this result would 
still hold if the magnetic charge of the monopole were ``hidden'' through 
spontaneous symmetry breaking.

This paper is dedicated to my grandparents Herbert and Annelise
Schmidt.


\begin{references}

\bibitem{dirac} P.A.M. Dirac (1931), Proc. Roy. Soc. {\bf A 133}, 
60; P.A.M. Dirac (1948), Phys. Rev. {\bf 74}, 817 

\bibitem{yang} T.T. Wu and C.N. Yang, Phys Rev. {\bf D12}, 
3845 (1975) 

\bibitem{cabbibo} N. Cabibbo and E. Ferrari, Nuovo Cimento 
{bf 23}, 1147 (1962)

\bibitem{zwang} D. Zwanziger, Phys. Rev. {\bf D3}, 880 (1971)

\bibitem{higgs} Higgs, P.W. (1964a) Phys. Letts. {\bf 12}, 132; 
Higgs, P.W. (1964b) Phys. Rev. Letts. {\bf 13}, 508; 
Higgs, P.W. (1966) Phys. Rev. {\bf 145}, 1156

\bibitem{jackson} J.D. Jackson, J.D. {\em Classical Electrodynamics} 
$2^{nd}$ Edition, (John Wiley \& Sons, 1975), p. 251

\bibitem{gold} H. Goldstein, {\em Classical Mechanics} $2^{nd}$
Edition, (Addison-Wesley Publishing Company, 1980), p.562

\bibitem{carmeli} M. Carmeli, {\em Classical Fields : General
Relativity and Gauge Theory} (John Wiley \& Sons, 1982), p. 590

\bibitem{salam} S.L. Glashow, Nucl. Phys. {\bf 22}, 579 (1961);
S. Weinberg, Phys Rev. Lett. {\bf 19}, 1264 (1967); Salam, A. 
{\em Elementary Particle Theory: Relativistic Groups
and Analiticity} (Nobel Symposium No. 8) edited by N. Svatholm,
1968 p. 367

\bibitem{saha} M.N. Saha, Ind. J. Phys. {\bf 10}, 145 (1936); 
M.N. Saha, Phys. Rev. {\bf 75}, 1968 (1949);H.A. Wilson, Phys. Rev. 
{\bf 75}, 309 (1949) 

\bibitem{goldhaber} A.S. Goldhaber, Phys. Rev. Letts. {\bf 36},
1122 (1976) 

\end{references}
\end{document}